\documentstyle[multicol,aps,epsf,amsmath]{revtex}
\begin{document}
\draft
\def\av#1{\langle#1\rangle}
\def\a{\alpha}
\def\l{\lambda}
\def\etal{{\it et al.}}
\def\pc{p_{\rm c}}
\def\df{d_{\rm f}}
\def\K{{\tilde K}}
\def\p{{\tilde p}}
\def\P{{\tilde P}}
\def\ie{{\it i.e.}}
\title {Scale-Free Networks are Ultrasmall}
\author{Reuven Cohen$^{1}$
\footnote  {{\bf e-mail:} cohenr@shoshi.ph.biu.ac.il}
 and Shlomo Havlin$^1$}
\address{$^1$Minerva Center and Department of Physics, Bar-Ilan university,
Ramat-Gan, Israel}
\maketitle
\begin{abstract}
We study the diameter, or the mean distance between sites, in a 
scale-free network, having $N$ sites and degree distribution
$p(k)\propto k^{-\l}$, {\it i.e.} the probability of having $k$ links outgoing
from a site. 
In contrast to the diameter of regular random networks or 
small world networks which is known 
to be $d\sim\ln N$, we show, using analytical arguments, that scale
free networks with $2<\l<3$ have a much smaller diameter, behaving as
$d\sim \ln\ln N$. 
For $\l=3$, our analysis yields $d\sim \ln N/\ln\ln N$, 
as obtained by Bollobas and Riordan, while for $\l>3$, $d\sim \ln N$.
We also show that, for any $\l>2$, one can construct a deterministic scale 
free network with $d\sim\ln\ln N$, and this construction yields the lowest
possible diameter.
\end{abstract}
\begin{multicols}{2}

It is well known \cite{bol,chung,bar_www,krapivsky} 
that random networks, such as Erd\H{o}s-R\'enyi networks~\cite{ER59,ER60} as 
well as partially random networks, such as small-world networks \cite{watts}, 
have a very small average distance (or diameter) between sites, 
which scales as $d\sim \ln N$, where $N$ is the number of sites. 
Since the diameter is small even for large $N$, it is common to
refer to such networks as ``small world'' networks.
Many natural and man-made networks have been shown to posses a scale
free degree distribution, including the Internet \cite{fal}, WWW 
\cite{bar_www,broder}, metabolic \cite{meta} and cellular
networks \cite{cell} and also trust cooperation networks \cite{amaral}. 

The question of the diameter of such networks is fundamental in the study of 
networks. It is relevant in many fields regarding communication and 
computer networks, such as routing~\cite{goh}, 
searching~\cite{search} and transport of information~\cite{goh}.
All those processes become more efficient when the diameter is smaller.
It also might be relevant to subjects such as the efficiency of chemical and
biochemical processes and spreading of viruses, rumors, etc. in cellular,
social and computer networks. In physics, the scaling of the 
diameter with the network size is related to the physical concept 
of the dimensionality of the system, and is highly
relevant to phenomena such as diffusion, conduction and transport. The 
anomalous scaling of the diameter in those networks is expected to lead 
to anomalies in diffusion and transport phenomena on those networks.
In this Letter we study the diameter of scale-free random networks and show 
that it is significantly smaller than the diameter of regular random networks.
We find that scale free networks with $2<\l<3$ have diameter $d\sim\ln\ln N$
and thus can be considered as ``ultra small world'' networks.

We define the diameter of a graph as the average distance
between any two sites on the graph (unlike the usual mathematical
definition
of the largest distance between two sites). Since no embedding space is defined
for those networks the distance denotes the shortest path between two sites
({\it i.e.} the smallest number of followed links needed to reach one from 
the other).
If the network is fragmented
we will only be interested in the diameter of the largest cluster (assuming
there is one). 

To estimate the diameter we will study the radius of such graphs.
We define the radius of a graph as the average distance of all sites on the 
graph from the site with the highest degree in the network (if there is 
more than one we will arbitrarily choose one of them). The diameter of the
graph, $d$, is restricted to:
\begin{equation}
r\leq d\leq 2r,
\end{equation}
where $r$ is the radius of the graph, defined as the average distance $\av{l}$
between the highest degree site (the origin) and all other sites.

A scale free graph is a graph having degree distribution, {\it i.e.} the 
probability that a site has $k$ connections:
\begin{equation}
p(k)=ck^{-\l}, \quad k=m,m+1,...,K,
\end{equation}
where $c\approx (\l-1)m^{\l-1}$ is a normalization factor, and $m$ and $K$ are
the lower and upper cutoffs of the distribution, respectively. The ensemble of
such graphs has been defined in \cite{aiello}. However, we will refer here to
the ensemble of scale free graphs with the ``natural'' cutoff
$K=mN^{1/(\lambda-1)}$ \cite{cohen,dor,rem1}.

We begin by showing that the lower bound on the diameter of any scale-free 
graph with $\l>2$ is of the order of $\ln \ln N$, then we show that for random
scale free graphs with $2<\l<3$ the diameter actually scales as $\ln\ln N$.  
It is easy to see that the lowest diameter for a graph with a given 
degree distribution is achieved by the following construction:
Start with the highest degree site, and then in each layer 
attach the next highest degree sites until the layer is full. By construction, 
loops will occur only in the last layer. This structure is somewhat similar to 
a graph with assortative mixing \cite{assort} -- since high degree sites tend
to connect to other high degree sites.

\begin{figure}
\narrowtext
\epsfxsize=2.5in
\hskip 0.3in
\epsfbox{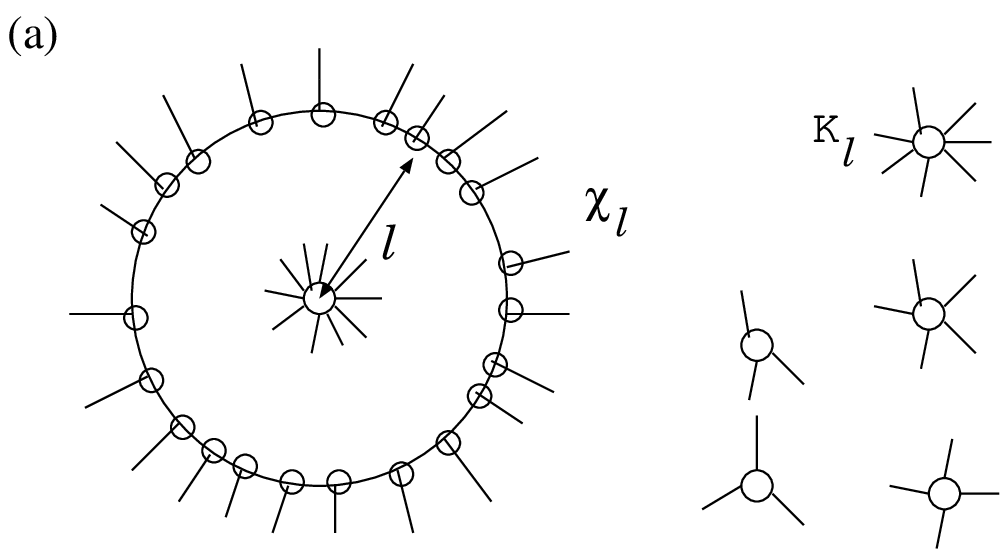}
\epsfxsize=2.5in
\vskip 0.15in
\hskip 0.3in
\epsfbox{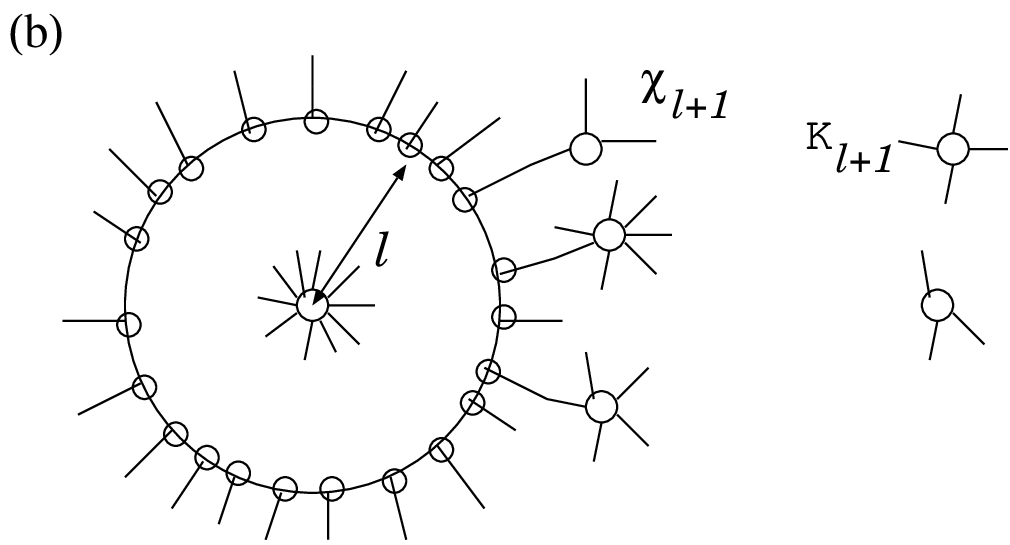}
\vskip 0.15in
\caption{Illustration of the exposure process. The large circle 
denotes the exposed fraction of
the giant component, while the small circles denote individual sites.
The sites on the right have not been reached yet.
(a) The structure after the exposure of the $l$th layer.
(b) The structure after the exposure of the $l+1$ layer.
\label{illust} }
\end{figure}

In this kind of graph the number of links outgoing from the $l$th 
layer (sites at distance $l$ from the origin), $\chi_l$, 
equals the total number of sites with degree between 
$K_l$, which is the highest degree of a site not reached in the $l$th
layer, and $K_{l+1}$, which is the same for the $l+1$ layer. See Fig.
\ref{illust}.
This can be described by the following equation:
\begin{equation}
\chi_l=N\int_{K_{l+1}}^{K_l} p(k)dk \approx 
m^{\l-1} N K_{l+1}^{1-\l}.
\end{equation}
The number of links outgoing from the $l+1$ layer equals the total number 
of links in all the sites between $K_l$ and $K_{l+1}$ minus one link at every
site, which is used to connect to the previous layer:
\begin{equation}
\chi_{l+1}=N\int_{K_{l+1}}^{K_l} (k-1)p(k)dk \approx \frac{\l-1}{\l-2}
m^{\l-1} N K_{l+1}^{2-\l}.
\end{equation}
Solving those recursion relations with the initial conditions 
$K_0=N^{1/(\l-1)}$ and $\chi_0=K_0$ leads to:
\begin{equation}
\chi_l=a^{(\l-1)(1-u^l)}N^{1-u^{l+1}},
\end{equation} 
where $a=(\l-1)/(\l-2)m$, $u=(\l-2)/(\l-1)$, and
\begin{equation}
\label{Kl1}
K_l=m (\chi_l/N)^\frac{1}{1-\l}\;.
\end{equation}

To bound the radius, $r$, of the graph, we will assume that the low degree 
sites are connected randomly to the center. We choose some degree 
$1\ll k^*\approx(\ln \ln N)^{1/(\l-1)}$. 
We can use Eq. (\ref{Kl1}) to show that if 
$l_1 \approx \ln \ln N/\ln (\l-2)$ then $K_{l_1}<k^*$, so all sites with 
degrees $k\geq k^*$ would have been reached with probability $1$ 
in the first $l$ layers. On the other hand, if we start uncovering
the graph from any site -- provided it belongs to the giant component --
then at a distance $l_2$ from this site there are at least $l_2$ bonds.
The probability that at none of those bonds will lead to a site of 
degree $k^*$ decays as $(1-k^* p(k^*)/\av{k})^{l_2}$. So, taking 
${k^*}^{\l-1}\ll l_2\ll\ln\ln N$, we will 
definitely reach a site of degree
at least $k^*$ at distance $l_2$ from almost every site. Since 
$l=l_1+l_2$, all those sites are at a distance of order $\ln \ln N$ 
from the highest degree site, this is the behavior of the radius 
of the graph. Thus, $\ln\ln N$ is a lower bound for the diameter of scale free
networks, and by applying this approach one can generate scale free 
networks with this diameter, for $\l>2$. For $\l=2$ the construction is 
somewhat similar to the condensate obtained in \cite{bianconi}. 
 
In the following, we present analytical arguments showing that the 
behavior of $d\sim \ln \ln N$ is actually
achieved in random non-correlated scale free graphs with $2<\l<3$.
Non-correlated networks are networks in which the degree of a site reached by
following a link is independent of the degree of the site at the other side of
that link, 
One can view the process of uncovering the network (which is the same as 
building it) by following the links one at a time. For simplicity let us 
start with the site with the highest degree 
(which is also guaranteed to belong to the giant component), 
whose degree is proportional to $N^{1/(\l-1)}$ \cite{cohen,dor}.
Next we expose the layers, $l=1,2,3,\ldots$, one
at a time. To this end we consider the graph as built from one large
developing cluster, and sites which have not yet been reached (they can also
belong to the giant component or not). A similar consideration has 
been used by Molloy and Reed \cite{molloy2}.

After layer $l$ is explored the distribution of the unreached sites
changes (since most high degree sites are reached in the first layers).
To take account for this we assume that
the $l$th layer has $\chi_l$ outgoing links. The distribution of
degrees after uncovering some of the edges changes to 
$P'(k)\approx P(k)exp(-k/K_l)$~\cite{molloy2}. In the limit of large $N$ and
large $K_l$ we will assume that after exploring this layer the highest 
degree of the unvisited sites 
is of order $K_l$, where $\chi_l$ and $K_l$ are functions 
of $l$ that will be determined later. 

Let us now consider the $l+1$ layer. There is a new
threshold function, that is, the new distribution of unvisited sites is like a
step function - almost $P(k)$ for $k<K_{l+1}$ and almost 
$0$ for $k>K_{l+1}$. The 
reason is as follows. A site 
with degree $k$ has a probability of $p=k/(N\av{k})$ to be reached by 
following a link \cite{rem2}. If there are  $\chi_l$ outgoing links 
then if $p\chi_l>1$ we 
can assume that (in the limit $N\to\infty$) the site will be reached 
in the next level with probability 1. Therefore, all unvisited 
sites with degree $k>N\av{k}/\chi_l$ will be surely 
reached in the next layer. On the other hand, almost all the 
unvisited sites with degree $k<N\av{k}/\chi_l$ will remain unvisited in 
the next layer - therefore, their distribution will remain almost
unchanged. From those considerations the highest degree of the 
unexplored sites in the $l+1$ layer is determined by:
\begin{equation}
\label{new_cutoff}
K_{l+1}\approx N\av{k}/\chi_l \;.
\end{equation} 

In the $l$th layer the number of loops, \ie the number of links connecting two
sites of the $l$th layer and the number of sites in the $l+1$ layer connected to
more than one site in the $l$th layer is proportional to $\chi_l^2/(\av{k}N)$.
Since as long as $\chi_l$ is not of order $N$, this fraction is smaller in order
than $\chi_l$, we can safely assume that loops can be neglected until the last
shells have been reached. Similar arguments have been used in \cite{cohen}.

In the $l+1$ layer, all sites with degree $k>N\av{k}/\chi_l$ will be 
exposed. Since the probability of reaching a site via a link is proportional
to $kP(k)$, the average degree of sites reached by following
a link is $\kappa\equiv \av{k^2}/\av{k}$~\cite{cohen}. 
For scale-free graphs $\kappa$ can be approximated by~\cite{cohen}
\begin{equation}
\label{kappa}
\kappa=\biggl(\frac{\l-2}{\l-3}\biggr)
\biggl(\frac{K^{3-\l}-m^{3-\l}}{K^{2-\l}-m^{2-\l}}\biggr)\;.
\end{equation}
This will be the average 
degree for sites reached in this layer, whose degree is 
$k<N\av{k}/\chi_l$. Therefore, $\kappa$ should be calculated using the new 
cutoff (\ref{new_cutoff}), from (\ref{kappa}) follows 
$\kappa\sim K_{l+1}^{3-\l}$.

Using the above considerations, the number of outgoing links from the $l+1$  
layer can be calculated.
To this end we consider the
total degree of all sites reached in the $l+1$ level. This includes
all sites with degree $k$, $K_{l+1}<k<K_l$, as well as other 
sites with average degree proportional to $\kappa-1$ links 
(the $-1$ is due to one link going inwards). Thus, the value of $\kappa$ is
calculated using the cutoff $K_{l+1}$. Loops within a layer and multiple links
connecting the same site in the $l+1$ layer can be neglected since as long 
as the number of sites in the layer are of order less than $N$, 
they are negligible in the limit $N\to\infty$.
The two contributions can be written as the sum of two terms:
\begin{equation}
\label{chi}
\chi_{l+1}=N\int_{K_{l+1}}^{K_l}(k-1)p(k)dk+
\chi_l\,[\kappa(K_{l+1})-1].
\end{equation}
Noting that $p(k)\propto k^{-\l}$ and that $\kappa\propto K^{3-\l}$
\cite{cohen} it follows that $\chi_{l+1}\propto NK_{l+1}^{2-\l}$ 
(where both terms in Eq. (\ref{chi}) 
contribute the same order). This can be written as a second
recurrence equation:
\begin{equation}
\label{rec_chi}
\chi_{l+1}=AN K_l^{2-\l}\;,
\end{equation}
where $A=\av{k}m^{\l-2}/(3-\l)=\frac{(\l-1)m}{(\l-2)(3-\l)}$. 

Solving the equations (\ref{new_cutoff}) and (\ref{rec_chi}) yields the result,
\begin{equation}
\chi_l\sim A^{\frac{(\l-2)^l-1}{\l-3}}N^{1-{\frac{(\l-2)^{l+1}}{\l-1}}},
\end{equation}
where $\chi_l$ is the number of outgoing links from the $l$th layer. Eq. 
(\ref{new_cutoff}) then leads to:
\begin{equation}
\label{Kl}
K_l\sim A^{\frac{(\l-2)^{l-1}-1}{3-\l}}N^{\frac{(\l-2)^l}{\l-1}}.
\end{equation}
Using the same considerations that follow Eq. (\ref{Kl1}), one can deduce that
here also 
\begin{equation}
\label{lnln}
d\approx \ln\ln N/\ln (\l-2).
\end{equation}
Our result, Eq. (\ref{lnln}), is consistent with the observations that 
the distance in the Internet network is extremely small and that the distance 
in metabolic scale free networks is almost independent of $N$ \cite{meta}. These
results can be explained by the fact that $\ln\ln N$ is almost a constant over
many orders of magnitude. Our
arguments show, however, that for a fixed distribution and very large values of
$N$ no scale-free graph with $\l>2$ can have a constant diameter. However, for
$\l=2$, since the highest degree site has order $N$ links, we expect that for
this case $d\approx const$.

For $\l>3$ and $N\gg 1$, $\kappa$ is independent  of $N$, and since the
second term of Eq. (\ref{chi}) is dominant, Eq. (\ref{chi}) 
reduces to $\chi_{l+1}=(\kappa-1)\chi_l$, 
where $\kappa$ 
is a constant depending only on $\l$. 
This leads to the known result $\chi_l\approx C(N,\l)(\kappa-1)^l$ and the 
radius of the network $l\propto \ln N$ \cite{newman}.

For $\l=3$, Eq. (\ref{chi}) reduces to $\chi_{l+1}=\chi_l\ln\chi_l$. Taking 
the logarithm of this equation one obtains 
$\ln\chi_{l+1}-\ln\chi_l=\ln\ln\chi_l$. Defining $g(l)=\ln\chi_l$ and 
approximating the difference equation by the differential equation $g'=\ln g$.
This equation can not be solved exactly. However, taking $u=\ln g$ the 
equation reduces to
\begin{equation}
\label{l3}
l=\int_{\ln\ln\sqrt{N}}^{\ln\ln N} e^{u-\ln u} du.
\end{equation}
The lower bound is obtained from the highest degree site for $\l=3$,
having degree $K=m\sqrt{N}$. Thus, $\chi_0=m\sqrt{N}$. The upper bound
is the result of searching $l$ for which $\chi_l\sim N$ with lower order
corrections.
The integral in Eq. (\ref{l3}) can be approximated by the steepest descent 
method, leading to
\begin{equation}
\label{rad3}
l\approx \ln N/(\ln\ln N),
\end{equation}
assuming $\ln\ln N\gg 1$. 

The above result, Eq. (\ref{rad3}), has been obtained rigorously for the 
maximum distance in the Barabasi-Albert (BA) model~\cite{bar}, 
having $\l=3$ (for $m\geq 2$) \cite{bol2}. Although the result in \cite{bol2}
is for the largest distance between two sites, their derivation makes it clear
that the average distance will also behave similarly. 
For $m=1$ in the Barabasi-Albert
model the graph turn into trees and the behavior of $d\sim \ln N$ is 
obtained \cite{bol2,szabo}. It should be noted that for $m=1$ the giant 
component in the random model contains only a fraction of the sites (while
for $m\geq 2$ it contains all the sites --- at least in the leading order). The
BA model, on the other hand, is fully connected for every $m$.
This might explain why exact trees and BA trees are different from generalized
random graphs.

Our derivation is valid for uncorrelated networks. For assortative networks
\cite{assort} the diameter is expected to be even smaller, as mentioned earlier.
For disassortative networks we would expect the odd layers to hold high degree
nodes and the even layers to hold low degree nodes, so it is plausible that 
the scaling of the diameter is the same, with some possible constant factor
$\leq2$. Note that this argument may not be valid for disassortative networks 
with $m=1$, where many dead-ends exist.

In summary, we have shown that scale free graphs have diameter 
$d\sim \ln\ln N$, which is smaller
than the $d\sim\ln N$ behavior, expected for regular random graphs.
For every $\l>2$ scale free graphs can be built to have a diameter
of order $d\sim\ln \ln N$. If random scale free graphs are considered, only
for $2<\l<3$, the behavior $d\sim\ln\ln N$ is obtained, while for $\l>3$ 
the usual result $d\sim\ln N$ is recovered. 

{\bf Note:}
After this manuscript \cite{cond_ver} has been submitted, two other 
manuscript have been submitted that confirm our results~\cite{DM02,CL02}.

\acknowledgments
We would like to thank Daniel ben-Avraham and Tomer Kalisky for useful 
discussion.

\end{multicols}

\begin{thebibliography}{99}
\bibitem{rem1}
   This cutoff is chosen since this is the ``natural'' cutoff of the network,
   due to the finite size of the system, if no external limitation is 
   imposed, see \cite{cohen,dor}. However, as can be seen from equations
   (\ref{new_cutoff}) and (\ref{rec_chi}),
   every cutoff above $\sqrt{N}$ will give similar results. For the
   deterministic graphs every power of $N$ will give similar results.
\bibitem{rem2}
   We assume that $\av{k}$ for the unvisited sites is fixed since it 
   is controlled by the low degree nodes - whose distribution is 
   unchanged.
\end{thebibliography}
\end{document}